
\input phyzzx

\def\plb{ Phys. Lett. }
\def\npb{ Nucl. Phys. }

\def\define#1#2\par{\def#1{\Ref#1{#2}\edef#1{\noexpand\refmark{#1}}}}
\def\con#1#2\noc{\let\?=\Ref\let\<=\refmark\let\Ref=\REFS\let\refmark=
\undefined#1\let\Ref=\REFSCON#2\let\Ref=\?\let\refmark=\<\refsend}

\define\BZJ
E. Br\'{e}zin and J. Zinn-Justin, \plb {\bf B288} (1992) 54.

\define\GZJ
P. Ginsparg and J. Zinn-Justin, "Action Principle and Large Order
Behavior of Non-Perturbative Gravity",  in {\it Random Surfaces and
Quantum Gravity}, Proc. of Carg\`{e}se workshop, May 28-June 1, 1990,
eds. O.Alvarez, E. Marinari and P. Windey.

\define\BX
L. Bonora and C.S. Xiong, SISSA preprint SISSA-92-161, hep-th/9209041,
(September, 1992); L. Bonora, M. Martellini and C.S. Xiong, \npb
{\bf B375}(1992)453.

\define\DSS
M. Douglas, N. Seiberg and S. Shenker, \plb {\bf B244} (1990) 381

\define\ME
H.B. Gao, preprint ZIMP 92-19 (September, 1992).

\define\K
K. Symanzik, Lett. Nouvo Cim. 6(2)(1973)77.

\define\GK
K. Gawedzki and A. Kupiainen, \npb {\bf B257[FS14]}(1985)474;
K. Gawedzki, A. Kupiainen and B. Tirozzi, \npb {\bf B257[FS14]}
(1985)610.

\titlepage
\hfill{
IC 302/92}

\hfill{
ZIMP 92-18}\break
\title{On Renormalization Group Flow In Matrix Model}
\author{ H.B. Gao\foot{Address after Oct. 1, Fakult\"{a}t f\"{u}r
Physik, Universit\"{a}t Freiburg, D-7800 Freiburg, Germany}}
\address{International Centre for Theoretical Physics,
P.O. Box 586, I-34100 Trieste, Italy}
\address{
and}
\address{Zhejiang Institute of Modern Physics, Zhejiang\foot{Permanent
address.} University,
Hangzhou 310027, Peoples Republic of China}
\abstract{The renormalization group flow recently found by Br\'{e}zin
and Zinn-Justin by integrating out redundant entries of the
 $(N+1)\times(N+1)$
Hermitian random matrix is studied. By introducing explicitly the
RG flow parameter, and adding suitable counter terms to
the matrix potential of the one matrix model, we deduce some
interesting properties of the RG trajectories. In particular, the
string equation for the general massive model interpolating between
the UV and IR fixed points turns out to be a consequence of RG
flow. An ambiguity in the UV
region of the RG trajectory is remarked to be related to the
large order behavior of the one matrix model.
}
\vfill
\endpage

Recently, Br\'{e}zin and Zinn-Justin\BZJ have suggested an approximate
scheme for studying the renormalization group (RG) flow in the one
matrix model. By treating the size $N$ of the hermitian matrix as an
effective cut-off in the theory, they have derived a differential
equation for matrix free energy, and reproduced approximately
the scaling law of the (multi-)criticality of the one matrix model.
Compare to the exact result of the double scaled limit,
their approach seems to be on the right track of an alternative
understanding \ME\ of the puzzles \GZJ\ related to the  KdV flows
in the one matrix model noted earlier.

Because of the well known problem of even potential matrix
integral, people hoped to be able to use the KdV flows
existed in the model to reach a well defined pure gravity theory
from the higher multicritical point. This has failed since the
flow itself suffers from instability problem \DSS\ .   Besides, we
lack an intrinsic way to see how the generalized KdV equations
and the general massive theory interpolating between
multicritical points are inter-related.

In this short note, we try to derive something interseting and
{\it exact} for the RG flows of Br\'{e}zin and Zinn-Justin.
The word exact will be explained later.
Note that eventhough, the values of the parameters at criticality
are calculated only approximately, the RG trajectory slightly
off the criticality may well survive  higher order corrections.

The starting point is the  potential of $N \times N$ hermitian matrix
for $m=2$ critical point (pure gravity) used by Br\'{e}zin and
Zinn-Justin:
$$V(\phi)=N({1 \over 2}tr\phi^2+{g \over 4}tr\phi^4). \eqno(1)$$
In [1] the following effective potential is obtained after
integrating out redundant entries of the $(N+1)\times(N+1)$
matrix,
$$V'(\phi)=(N+1)({1 \over 2}tr\phi^2+{g \over 4}tr\phi^4)
+gtr\phi^2, \eqno(2)$$
where, in both equations (1)and (2), $\phi$ denotes
$N \times N$ matrix. In deriving (2), a parametrization of matrix
$$\phi_{(N+1)\times(N+1)}=\pmatrix{\phi_{N \times N} & {\bf u} \cr
\bar{\bf u} & \alpha \cr} \eqno(3)$$
is used, and the assumption is made setting $\alpha=0$.

In the sense of a RG flow from $N$ to $N+1$, equation (2)
corresponds to adding a counter term $(1/2 +g)tr\phi^2+g/4tr\phi^4$
to the original potential (1). We observe that the parameter
of RG transformation, $t\sim {1 \over N}$, is implied implicitly (which
might have shown up in the rescaling $N\rightarrow N'=e^t N$).
Since in the large $N$ limit, the model is driven to an (IR) fixed
point which describes pure gravity, along the trajectory of increasing
$t$, quantities of order $ { 1 \over N}$ which were previously ignored
become important in a hypothetic massive theory. The question is then
what have been thrown away before? Obviously, $\alpha \sim {1 \over N}$
and which has been set to zero when deriving (2). Our assumption here
is that a non-zero $\alpha$ in (3) substitutes for the role of the
implicit parameter $t$ of the RG transformation induced by integrating
the margins of the $(N+1)$-matrix. Thus we modify the RG flow equation
so that it is
satisfied with a small, un-integrated parameter $\alpha$.

The above assumption leads, by a calculation similar to that in [1],
to the effective potential
$$V''(\phi)=(N+1)({1 \over 2}tr\phi^2 +{g \over 4}tr\phi^4)
+gtr\phi^2 + \gamma tr\phi, \eqno(4)$$
where, $\gamma=\alpha g$, and only terms of first order in $\alpha$
are kept. It is easy to check that (4) does not alter the (IR)
fixed point structure as long as one also adds a counter term
proportional to $tr\phi^3$.

Therefore, in the same sense as introducing counter terms
corresponding to (2) does not alter the fixed point structure of (1),
further counter terms can be added
$$\gamma tr\phi + \lambda tr\phi^3  \eqno(5)$$
without violating the original scaling law of (1), at the present
level of accuracy.

Note that the condition of tadpole cancellation imposes a
relation between the two coefficients in (5) whose explicit form
we will not write. This implies that when $\alpha$ is interpreted
as RG flow parameter $t$, $\lambda (t)$ evolves in the augmented
space of coupling constants $(g, \lambda)$.

The generalization to the multi-critical points is straightforward
and the result is a structure of counter terms consisting
of traces of all odd powers of $\phi$,
$$ \sum \lambda_{2k+1}
tr\phi^{2k+1}, ~~~~~~ k=0, 1, 2... \eqno(6)$$

To recapitulate, we have assumed a non-zero $\alpha$ which plays
the role of RG flow parameter, and obtained a doubling of matrix
model couplings. The appearance of terms involving odd power of
$\phi$ should be interpreted as turning on relevant perturbations.

The immediate consequence of the doubling of couplings is the
possiblity of extracting massive model from the RG flow of the
critical point
models. In fact, starting from a general matrix model potential
$$V(\phi)=\sum_{k=1} t_k tr\phi^k, ~~~~~ k=1,2,...\eqno(7)$$
and defining an {\it abstract} linear system \BX\ , one is able
to "derive" the following form of string equation for general
massive model connecting  multicritical points in
one matrix model
$$\sum_{m=1}m \tilde{t}_m {\partial \over \partial
\tilde{t}_{m-1}}F- { \partial F \over \partial \tilde{t}_0}
+ {1 \over 2} \tilde{t}^2_0 =0, \eqno(8)$$
where $F$ is the string partition function and the flow times
$\tilde{t}_m$ and the matrix couplings $t_k$ are related by
$\tilde{t}_m \sim t_{2k+1}$ upto a suitable rescaling.

A few words about the exactness of our result. Note that the
most natural interpretation of the string equation (8) is
in terms of a special discrete linear system \BX\ , of which
the string equation is a  consequence of compatibility
condition. Equation (8) looks so familiar to us that it
should survive the {\it continuum } limit, although this
limit is not at all necessary \BX\ as long as the topological
nature is concerned. It is in this sense that we have deduced
exact result through approximate RG transformation. At
any rate, a massive theory itself is not responsible for
its crude asymptotics (but the boundary conditions are).

We conclude this letter with a comment on the possible cure of
the problem associated to the KdV flows. The RG flow of [1]
is formally asymptotic to the IR regime. Because we have
made use of the fact $\alpha$ small, formal procedure for
studying the $UV$ asymptotics is  broken down. This is also to
be attributed to some sort of instabilty problem, but this time
in the $UV$ regime. However, a trick \K\ which was used in
studying the $\lambda \phi^4$ field theory could be of help
here. That is to analytically continue to the negative
coupling in such a way the UV regime is approachable from
the negative side near the $zero$ coupling point \GK\ .
We have carried out a study along this line and indeed
found a flow function near the UV regime which is free of
the Borel/instanton singularities \ME\ .

\centerline{\bf{Acknowledgements}}

I thank Dr C.S. Xiong for an informal discussion on the results of
[5]. The author would like to thank the International Centre for
Theoretical Physics, Trieste for hospitality extended to him
when the work is completed. This work is supported in part by
the Zhejiang Provincial NSF.

\refout
\end